\documentclass[aps,pra,showpacs,twocolumn,superscriptaddress]{revtex4}
\usepackage{graphicx}
\usepackage{amsmath,amssymb,amsthm,dsfont,bm}

\begin{document}
\preprint{}
\title{Nonlinear Michelson interferometer for improved quantum metrology}
\author{Alfredo Luis}
\email{alluis@fis.ucm.es}
\homepage{http://www.ucm.es/info/gioq}
\affiliation{Departamento de \'{O}ptica, Facultad de Ciencias
F\'{\i}sicas, Universidad Complutense, 28040 Madrid, Spain}
\author{\'{A}ngel Rivas}
\affiliation{Departamento de F\'isica Te\'orica I, Facultad de Ciencias F\'isicas,
Universidad Complutense, 28040 Madrid, Spain}

\date{\today}

\begin{abstract}
We examine quantum detection via a Michelson interferometer embedded
in a gas with Kerr nonlinearity. This nonlinear interferometer is illuminated by pulses of classical
light. This strategy combines the robustness against practical imperfections of classical
light with the improvement provided by nonlinear processes. Regarding ultimate quantum
limits, we stress that, as a difference with linear schemes, the nonlinearity introduces
pulse duration as a new variable into play along with the energy resources.
\end{abstract}

\maketitle

\section{Introduction}

Precise measurements are crucial in physics since they constitute the
link between the theory and nature. Accurate measurements can promote
or reject a theory. Besides, precise detection and monitoring are
fundamental for technology and other applications of science.

A critical  contribution of the quantum theory to metrology is that quantum
fluctuations would limit the resolution to some ultimate limits depending
on the energy resources employed in the process \cite{GLM11}, usually
counted as the number of particles.

Since standard metrology is based on linear processes, previously
known quantum limits heavily depend on an implicit assumption of
linearity. Thus, a new frontier arises if we consider that the signal
may be detected via nonlinear processes. The key point is that nonlinear
schemes allow us to reach larger resolution than linear ones for the same
resources. Moreover, the improvement holds even when using probes in
classical states. This is of much relevance concerning robustness against
practical imperfections, which can be deadly
for schemes based on nonclassical probe states \cite{HMPEPC97,DBS13,DJK14}.
Quantum nonlinear metrology has been studied and proven experimentally
in very different physical contexts
 \cite{LU04,BFCG07,BDDSTC09,NM10,TBDSC10,RL10,LU10,NKDBSM11,SNBCCM14}.
In particular, this is the case for light propagation in Kerr-type nonlinear
media, that has already demonstrated  its usefulness in the context of
precise detection \cite{PCW93,RHSD95,KTRLWDS09}.

In this paper we present a new feature of quantum detection involving nonlinear processes.
This is that resolution depends not only  on the number of photons but
also on the duration of the pulse (this is both on the number of
particles and on the rate at which the are employed). This is in sharp contrast
to linear schemes where the duration of the pulse plays no role. In this
way a new variable appears which may be advantageously used to
improve detection performance beyond previously accepted limits.

\section{Scheme}

\begin{figure}
\begin{center}
\includegraphics[width=6cm]{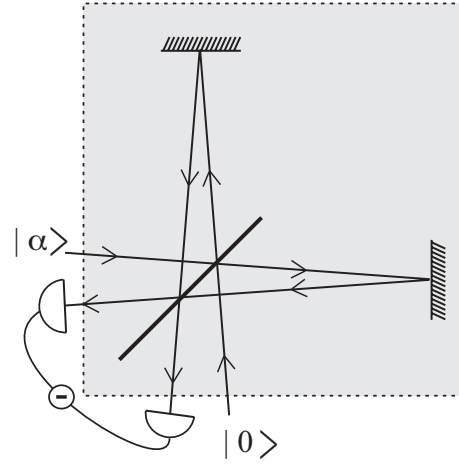}
\end{center}
\caption{Scheme of a Michelson interferometer embedded in a nonlinear medium.}
\end{figure}

To develop this point let us consider signals encoded as  length variations that can
be typically detected via a Michelson interferometer schematized in Fig. 1. In the absence
of a signal the two arms are equal, $\ell_2 = \ell_1 = \ell_0$. For simplicity we will
consider the signal manifests as an anti-correlated length change as $ \ell_1 =
\ell_0 - x/2$, $\ell_2 = \ell_0 + x/2$ as it is expected to be the case by the pass of a
gravitational wave \cite{AD14}.

In order to involve nonlinear effects, we assume that the interferometer is embedded
in a gas displaying a Kerr nonlinearity. The light is made of classical-light pulses of
mean frequency $\omega$, duration $\tau$, and cross section $A$. The propagation
in the nonlinear medium can be conveniently expressed in terms of an intensity-dependent
index of refraction,
\begin{equation}
\label{nIN}
n = n_0 + \tilde{n} I = n_0 \left (
1 + \chi N \right ) ,
\end{equation}
here, $\tilde{n}$ is the nonlinear coefficient,  $n_0$ is the linear index, $I$ is the
light intensity, $N$ is the number of photons of each pulse, and $\chi$ expresses
the nonlinear phase shift per photon,
\begin{equation}
\label{chi}
I \simeq   \frac{\hbar \omega}{ A \tau} N , \quad \chi =  \frac{\tilde{n}}{n_0}
\frac{\hbar \omega}{ A \tau} ,
\end{equation}
where this is a definition of $\chi$ whereas the $I$ versus $N$ relation is an estimation
good enough for our purposes as far as the exact equivalence would require the
specific spectral distribution of the pulse.

After Eq. (\ref{nIN}), the light propagation within the interferometer is described in the
quantum domain by the unitary operator $U=U_1 U_2$ with
\begin{equation}
\label{generator}
U_j = e^{i \varphi_j G_j}, \qquad G_j = \hat{N}_j +
\frac{\chi}{2 } \hat{N}_j^2  ,
\end{equation}
with $\varphi_j = k \ell_j$  where $k= n_0 \omega /c$ is the wave number, and $\hat{N}_j=
a_j^\dagger a_j$  represents the photon-number operator  in each arm $j=1,2$. We further
assume that the  signal induces an arm length difference $x$ small compared with the
length of the pulse $c \tau$. Any other changes produced in the optical constants of the
media are assumed to lead to negligible effects.

We illuminate the interferometer just by one of the input ports (as usual the vacuum
is at the other input)  by a classical-like pure coherent state $| \alpha \rangle$ with a
mean number of photons $| \alpha |^2 = N \gg 1$ [exactly the same results are obtained
if the probe state before the input beam splitter is in the phase-averaged mixed
state $\int_0^{2\pi} d \theta | \alpha e^{i\theta} \rangle \langle  \alpha e^{i\theta} |/(2 \pi)$].
The light state in the internal modes $a_{1,2}$ of the interferometer leaving the 50\% beam
splitter can be expressed as the product of coherent states  $|\alpha/\sqrt{2} \rangle_1
|\alpha/\sqrt{2} \rangle_2$, whereas  the light state reaching the beam splitter after propagation
within the interferometer is $U_1 U_2 |\alpha/\sqrt{2} \rangle_1 |\alpha/\sqrt{2} \rangle_2$.

The measurement is carried out by registering the difference in the number of photons recorded
by two detectors at the output ports of the interferometer. The corresponding operator can be
expressed in terms of the internal modes of the interferometer $a_{1,2}$ as $M= i(a^\dagger_2  a_1 -
a^\dagger_1 a_2)$.

The sources of technical noise will be taken into account by their most typical consequences
such as phase randomization (e.g., caused by fluctuations in the optical properties of the medium),
thermalization (e.g., thermal photons coming from undesired residual sources), as well as the
usual finite quantum efficiency of the detectors.

\section{Signal detection and uncertainty}

The signal $x$ produces a phase shift that alters the statistics
of the observed $M$, shifting its mean value (see the Appendix for details),
\begin{equation}
\langle M \rangle = \eta N e^{-N\chi^2 k^2 x^2 /8} e^{-\sigma^2 /2} \sin \left [ kx \left (
1 +  \chi  \frac{N}{2} \right ) \right ] ,
\end{equation}
where $\eta$ is the quantum efficiency of the detectors and $\sigma$ is the variance
of the random  relative phase.

In order to be detected, the signal-induced shift  $\langle M \rangle$ must be larger
than the background quantum noise $\Delta M$ at $x =0$. Taking into account the
noise sources commented on above we get
\begin{equation}
\left ( \Delta M \right )^2 =\eta N + \eta^2 N^2 \sigma^2 +
\eta N N_t ,
\end{equation}
where $N_t$ is the mean number of the undesired thermal photons.
Since the nonlinearity complicates the noise analysis we postpone the
details of the calculation to an Appendix to not interrupt the analysis.

We can estimate the resolution of the detection of $x$ via the noise-to-signal ratio as
\begin{equation}
\label{snr}
\left ( \Delta x \right )^2 =
\frac{\left ( \Delta M \right )^2}{\left |
\frac{d \langle M  \rangle}{d x} \right |^2}   ,
\end{equation}
leading to
\begin{equation}
\label{Dx}
\left ( \Delta x \right )^2 =
\frac{1+ \eta N \sigma^2 + N_t }{\eta k^2  N \left ( 1 + \chi N/2 \right )^2}  .
\end{equation}
This holds provided that $\chi N k x \ll 1$, $N_t \ll N$, and $\sigma \ll 1$. Moreover,
we have assumed $\chi k \ell_0 \simeq 2 \pi m$  for integer $m$. Otherwise, without
this last condition the nonlinearity would deeply disturb the probe state by producing
coherent superpositions of distinguishable states \cite{YS86}, and the standard interferometric
measurement $M$ would become useless requiring  more advanced detection strategies
beyond the scope of the present analysis  (see the Appendix for further details). The nonlinearity
will have a noticeable effect for $\chi N \gg1$,  which is compatible with the
above assumption  $\chi N k x \ll 1$ provided that $kx$ is small enough $kx \ll1$.

This might be compared with the case when the nonlinear medium is absent $\chi =0$
\begin{equation}
 \left ( \left . \Delta x \right |_{\mathrm{lin}} \right )^2 = \frac{1+   N_t}{\eta  k^2 N} ,
\end{equation}
and we have further assumed that in such a case propagation occurs in vacuum and the
phase randomization can be safely neglected $\sigma=0$.

In the ideal case that the phase  randomization and thermal effects might be ignored
$\sigma = N_t =0$ and $\chi N \gg 1$ we get the following improvement of the nonlinear
versus the linear scheme,
\begin{equation}
\left ( \Delta x  \right )^2 = \frac{4 \left ( \left . \Delta x \right |_{\mathrm{lin}} \right )^2}{\chi^2 N^2 }
\rightarrow  \frac{4}{\eta k^2 \chi^2 N^3} .
\end{equation}
We recall that the duration of the pulses is embedded in the nonlinear phase shift per photon
$\chi$ in Eq. (\ref{chi}), and so, the lesser $\tau$, the larger $\chi$, and the larger the resolution.

\section{Discussion}

We can roughly estimate the amount of noise reduction with parameters within the
reach of current technology. For the sake of simplicity and to fix the main ideas
let us first consider the ideal case where the effect of technical noise is negligible
$\eta N \sigma^2 +  N_t \ll  1$ and $\eta \simeq 1$. Regarding numerical values we
can address two extreme situations: standard natural nonlinearities, and giant
nonlinearities achieved via atomic coherence. Throughout we will assume that the
index in darkness is on the order of unity $n_0 \simeq 1$.

The typical natural nonlinearities in gases can be on the order of $\tilde{n} \simeq
10^{-17} \textrm{cm}^2 / \textrm{W}$ \cite{BHKKO10}. As to the pulse parameters let
us assume a pulse duration of $\tau \simeq 1$ ps, light power of $P \simeq 1$ PW,
and beam cross section of $A \simeq 10^{-9} \mathrm{m}^2$,
which leads in the visible spectrum to $N \simeq 10^{21}$ photons per pulse and
a nonlinear shift per photon of $\chi \simeq 10^{-18}$, so that
\begin{equation}
\Delta x \simeq 10^{-3} \left . \Delta x \right |_{\mathrm{lin}} \simeq 10^{-21} \mathrm{m} .
\end{equation}
The  condition $\chi N k x \ll 1$ means the following condition on the signal
$x \ll 10^{-10} \textrm{m}$. On the other hand, the condition  $\chi k \ell_0 \simeq 2 \pi m$
leads to an extremely large interferometer even for $m=1$ since in such a case
$\ell_0  \simeq 10^{12} \mathrm{m}$. Thus the $m=0$ situation should be addressed as
suggested in the Appendix.

Things are completely different if we consider the giant nonlinearities achieved via
electromagnetically induced transparency, leading to Kerr coefficients on the order
of $\tilde{n} \simeq 10^{-2} \textrm{cm}^2 / \textrm{W}$  as reported in Ref. \cite{SRKD08}
and similarly large values in other configurations, such as  $\tilde{n} \simeq
10^{-5} \textrm{cm}^2 / \textrm{W}$ in Ref. \cite{FEFSPV08}. Such large values allows for
alleviating the requirements on the light probe state. For example we may have
$\tau \simeq 100$ ps, $P \simeq 1$ MW, and $A \simeq 10^{-6} \mathrm{m}^2$,
which leads in the visible spectrum to $N \simeq 10^{14}$ and $\chi \simeq 10^{-8}$
so that
\begin{equation}
\Delta x \simeq 10^{-6} \left . \Delta x \right |_{\mathrm{lin}} \simeq 10^{-20} \mathrm{m} .
\end{equation}
The condition  $\chi k \ell_0 \simeq 2 \pi m$ leads to a more practical interferometer with $\ell_0
\simeq 100\ \mathrm{m}$, whereas  the  condition $\chi N k x \ll 1$ implies $x \ll 10^{-13} \textrm{m}$.
Thus a detectable signal should be composed in the range $10^{-13} \mathrm{m} \gg x \gg
10^{-20} \mathrm{m}$. This fits perfectly well with the expected signals due to the pass of
a gravitational wave in a 100-m-long interferometer, which are  $10^{-15} \mathrm{m} \gg x \gg
10^{-20} \mathrm{m}$ \cite{RO00}. Notably, smaller $\tau$ and/or $A$, such as the
beam-size values reached in Ref. \cite{DelRe15} with current technology, may lead even to
room-size interferometers with similar performance.

Finally, we may estimate the maximum effect of imperfections so that the good effects of
nonlinearity are not spoiled. The condition we are looking for is derived from $\eta N
\sigma^2 + N_t  \ll \chi^2 N^2$ that is satisfied if, roughly speaking, $\sigma \ll \chi \sqrt{N/\eta}$
and $N_t \ll \chi^2 N^2$. For the natural nonlinearity and $\eta \simeq 1$ we get $\sigma \ll 10^{-8}$,
$N_t \ll 10^6$,  whereas for giant nonlinearities we get much less limiting bounds, $\sigma \ll 10^{-1} $,
$N_t \ll 10^{12}$.

\section{Conclusions}

Summarizing, nonlinearity not only can improve resolution beyond linear limits, but also introduces
a new variable into play. The signal uncertainty depends not only on the number of probe photons
$N$, but also on the duration of the pulse $\tau$ through the nonlinear effect per photon $\chi$ in
Eq. (\ref{chi}). This is because optical nonlinearity is sensible to light intensity rather than just energy
or photon number. In particular, after Eq. (\ref{Dx}) we may conjecture  an optimum ultimate quantum
limit (that would require nonclassical probes to be reached) scaling as
\begin{equation}
\Delta x \propto \frac{\tau A \lambda^2}{N^2} ,
\end{equation}
in terms of the probe free parameters, where $\lambda$ is the wavelength. This result may be
particularly useful for example in situations of frequent monitoring where small pulse durations
and large repetition rate of the interrogating pulse may be of interest. In this regard, the availability
to obtain large beam intensities by shortening pulses seems a more feasible condition than
increasing energy resources as required in usual linear quantum metrology.

\bigskip

We acknowledge financial support from Spanish Ministerio de Econom\'ia y Competitividad Projects No. FIS2012-33152 and No. FIS2012-35583 and from the Comunidad Autonoma de Madrid research consortium QUITEMAD+ Grant No. 210 S2013/ICE-2801.

\appendix

\smallskip

\section{Calculus}

After the relation $a F(a^\dagger a) = F(a^\dagger a+1) a$ valid for any $F$ we get that
\begin{equation}
U^\dagger a_j U = e^{i \varphi_j} e^{i z_j } e^{i 2 z_j \hat{N}_j}  a_j ,
\end{equation}
where $\varphi_j = k \ell_j$, $z_j = \varphi_j  \chi /2 $, and $U=U_1 U_2$ is the global
transformation.  When evaluating the mean values of  $a^\dagger_1 a_2$ and its Hermitian
conjugate on coherent states $| \beta \rangle$   with $\beta = \alpha /\sqrt{2}$, we will get
expressions of the form
\begin{equation}
\langle \beta | U^\dagger a_j U | \beta \rangle = e^{i \varphi_j} e^{i z_j }
\langle \beta | e^{i 2 z_j \hat{N}_j}  a_j | \beta \rangle  ,
\end{equation}
that can be easily evaluated taking into account that
\begin{equation}
\langle \beta | e^{i 2 z_j \hat{N}_j}  a_j | \beta \rangle = \beta \langle \beta  |  \beta e^{i 2 z_j }
\rangle = \beta e^{| \beta|^2 \left ( e^{i 2 z_j }-1 \right )} .
 \end{equation}

Besides the finite quantum efficiency of the detectors we will consider some further
typical forms of practical noise, such as thermalization and phase randomization. These
common noise forms can have different physical origins, such as  fluctuations of the
optical properties  of the medium, random variations of the complex amplitude from
pulse to pulse, and so on. They can be addressed at once by performing the
replacements,
 \begin{equation}
 \label{tf}
a^\dagger_1 a_2 \rightarrow e^{i \phi} \left (\sqrt{\eta} a^\dagger_1 +  \sqrt{1- \eta} \,
b^\dagger_1 \right ) \left (  \sqrt{\eta} a_2 + \sqrt{1- \eta} \, b_2 \right ) ,
 \end{equation}
where $\phi$ is a random phase that we will assume to be Gaussian distributed with zero mean
and variance $\sigma^2$, $\eta$ is the quantum efficiency in the detection, and $b_j$'s are
uncorrelated field modes in thermal states  with  $(1-\eta) \langle b^\dagger_1 b_1 \rangle =
(1-\eta)  \langle b^\dagger_2 b_2 \rangle = N_t /2$ with $N_t  \ll N$, and $ \langle b_1 \rangle = \langle b_2
 \rangle = \langle b^\dagger_1 b_2 \rangle =0$.

\subsection{Mean value}

Taking all this into account the mean value of $M$ can be obtained after a long but straightforward
calculation as

\begin{widetext}

\begin{equation}
\label{wt}
\langle M \rangle= \eta N e^{\frac{1}{2} N \left [ \cos \left ( 2 z_1 \right ) + \cos \left ( 2 z_2 \right )-
2 \right ] }   \sin \left \{ \phi + \varphi_2 - \varphi_1 + z_2 - z_1 + \frac{N}{2} \left [  \sin \left ( 2 z_2 \right ) -
\sin \left (2 z_1 \right ) \right ] \right \} .
\end{equation}

\end{widetext}

Before the $\phi$ integration several natural considerations  seem in order to get simpler and
meaningful expressions. A required condition is that the factor in the real exponential should be
close to zero, otherwise the final uncertainty $\Delta x$ would increase exponentially with $N$.
This is because the uncertainty $\Delta M$ will contain always a photon-counting noise term
independent of the arm lengths. Thus we have to consider that in the absence of signal $z_0 = m \pi$,
where $m$ is any integer. This may be achieved by properly adjusting the fixed arm length $\ell_0$
depending on $\chi$. Alternatively we may consider that the Kerr transformation induced by the
fixed length $\ell_0$ may be compensated by another Kerr transformation with a nonlinear susceptibility
of the opposite sign in propagation conditions insensitive to the signal value $x$. An alternative
approach may follow the strategy in Ref. \cite{JLZLBBNDW13} by comparing outputs for two
consecutive pulses experiencing alternatively linear and nonlinear transformations.

Thus, considering that the signal induces a very small variation in $z_j$ around $z_0 \simeq  m\pi$
 we have
\begin{equation}
\langle M \rangle \simeq   \eta  N e^{-  \chi^2 N k^2x^2 /8} \sin \left [ \phi + k x \left (1+
\chi \frac{N}{2}  \right)  \right ]  ,
\end{equation}
which can be obtained after Eq. (\ref{wt}) by a series expansion of the harmonic functions
within the exponential and the sine function where we have also neglected the $z_2 - z_1$
term not multiplied by $N$. Carrying out the $\phi$ integration over a Gaussian distribution
with zero mean and variance $\sigma^2$ we get
\begin{equation}
\label{M1}
\langle M \rangle \simeq  \eta  N e^{-  \chi^2 N k^2x^2 /8} e^{-\sigma^2 /2}
\sin \left [ k x \left (1+ \chi \frac{N}{2}  \right)  \right ]  .
\end{equation}
Finally, considering typical values for the variables involved the real exponentials can be safely
approximated by unity. Moreover, the expected signals are small enough $\chi N k x \ll 1$ so that
there is a linear relationship between $M$ and  $x$, which is usually an implicit assumption leading
to Eq. (\ref{snr}),
\begin{equation}
\langle M \rangle \simeq  \eta  N \left [ k x \left (1+ \chi \frac{N}{2}  \right)  \right ]  .
\end{equation}
This is compatible with the fact that $\chi N$  can be very large having observable effects.
This is because  it determines the value of the $x$ derivative in the denominator of Eq. (\ref{snr}) and such
a derivative need not be small. On the contrary, the best situation holds when $\chi N$ is large enough to imply
a noticeable reduction in signal uncertainty as analyzed in detail in Secs. III and IV.

\subsection{Uncertainty}

Next we address the evaluation of $(\Delta M )^2$  at $\ell_2 = \ell_1 = \ell_0$,  this is $x=0$ so
that  $\langle M \rangle = 0$ and $(\Delta M )^2 = \langle M^2 \rangle$. After Eq. (\ref{tf}) the effect
of thermalization and finite efficiency  means that $M^2$ should be replaced by
\begin{equation}
\label{tq}
\eta^2 M_0^2 + \eta (1 - \eta) \left [ \left ( 2 b^\dagger_2 b_2 + 1 \right ) \hat{N}_1
+ \left ( 2 b^\dagger_1 b_1 + 1 \right ) \hat{N}_2  \right ] ,
\end{equation}
where to avoid confusion we denote by $M_0$ when evaluating $M$ in the noiseless case.
Other terms lead to null or negligible contributions. It is worth noting that the last term is
not affected by the $U_j$  transformation nor by the random phase.

Then we can compute $\langle M_0^2 \rangle$, where
\begin{equation}
M_0^2 = 2 \hat{N}_1  \hat{N}_2 + \hat{N}_1  + \hat{N}_2- a^{\dagger 2}_1 a_2^2 - a^{\dagger 2}_2 a_1^2 .
\end{equation}
The first terms depending just on $ \hat{N}_j $ are invariant under the transformations $U_j$ whereas for
the remaining two terms we can use the fact that $a^2 F(a^\dagger a) = F(a^\dagger a+2) a^2$ and then proceed
as above, to get, before phase randomization,
\begin{equation}
\label{M2}
\langle M_0^2 \rangle= \frac{N^2}{2} + N - \frac{N^2}{2}  \cos ( 2 \phi) ,
\end{equation}
and after the random-phase average,
\begin{equation}
\label{pa}
\langle M_0^2 \rangle= \frac{N^2}{2} + N - \frac{N^2}{2}  e^{- 2 \sigma^2}  \simeq N + \sigma^2 N^2 ,
\end{equation}
where the approximation holds for $\sigma \ll 1$. Finally, collecting the contributions in Eqs. (\ref{tq}) and
(\ref{pa}) we finally get
\begin{equation}
\left ( \Delta M \right )^2 \simeq \eta N + \eta^2 \sigma^2 N^2 + \eta N N_t ,
\end{equation}
leading to Eq. (\ref{Dx}).

\end{document}